\documentclass[runningheads]{svmult}

\usepackage{makeidx}   
\usepackage{graphicx}  
\usepackage{subeqnar}  
\usepackage{multicol}  
\usepackage{physprbb}  
\makeindex             

\begin{document}
\title*{The GRB followup Euro-US Consortium:\protect\newline
results from the ESO telescopes}

\toctitle{The GRB followup Euro-US Consortium: results from the ESO
telescopes}

\titlerunning{The GRB followup Euro-US Consortium: results from the ESO
telescopes}

\author{Nicola Masetti\inst{1},
on behalf of a large collaboration}

\authorrunning{N. Masetti}

\institute{$^1$Istituto TeSRE, CNR, via Gobetti 101, I-40129 Bologna, Italy}

\maketitle              

\vspace{-.5cm}

\begin{abstract}
In October 1997, the Italian and Dutch GRB teams started a collaboration
on ESO optical follow-up of rapidly and accurately localized GRBs.
Subsequently, starting April 1, 2000, this collaboration was extended
to astronomers from other countries,
who contributed their expertise for the creation of a Consortium committed
to the study of GRB counterparts and host galaxies at optical and
near-infrared wavelengths.
The collaboration aims at the joint exploitation of the observations
taken within an ESO Large Programme approved for the two-year period
April 1, 2000 - March 31, 2002. Here we describe history and organization
of this Consortium, the goals of the ESO Large Programme, and the main
results obtained up to now with ESO telescopes.
\end{abstract}

\vspace{-.2cm}

\section{Introduction and History}

\vspace{-.2cm}

The availability of fast (few hours after the GRB trigger) and precise
(arcmin-sized) GRB localizations, first afforded by the Italian-Dutch
X--ray satellite {\it BeppoSAX} in 1996 and subsequently by other
spacecraft, allowed astronomers to pinpoint Optical
Transients (OTs) associated with the high energy events and to
effectively explore the physics behind these phenomena. In this
contribution we briefly outline history, status and results of the search
and followup of GRB optical and near-infrared (NIR) counterparts at ESO
since the beginning of the {\it BeppoSAX} afterglow era. 

The first ESO ToO observation of a fast and precise GRB position was
activated on January 1997 at NTT on the GRB970111 error box, but no
optical counterpart was detected.
The first detection of an OT associated with a GRB was achieved
at ESO about two months later, on 1997 March 13, with NTT observations of
the GRB970228 error box \cite{jvp}, the first GRB for which X--ray and
optical afterglows were discovered. 

Subsequently, thrusted by this outstanding scientific achievement, the
Italian and Dutch GRB search and followup teams, led by Filippo Frontera
and Jan van Paradijs, respectively, independently submitted regular ESO
proposals for Period 60 (October 1997 - March 1998) for the activation of
ToO observations at several ESO telescopes on GRB error boxes observable
from Chile. Both proposals were approved and, following the suggestion
from ESO, the Italian and Dutch groups decided to start collaborating in
this search and to form a single team. 

The Italian-Dutch collaboration in the GRB follow-up at the ESO telescopes
was organized in a way that the two groups alternated in the program lead
at every trigger of an accurately and rapidly localized GRB, independent
of spacecraft. The leadership of the group `on duty' encompassed every
step from alerting the telescopes to taking responsibility of results
publication.

Besides the Italian-Dutch collaboration, a parallel proposal set up
by Danish, Spanish and German teams was active at several ESO telescopes.

During Periods 60 to 64 (October 1997 - March 2000)
ESO observations placed several milestones in the
study of GRB optical afterglows, in particular: the detection of SN 1998bw
in the error box of GRB980425, the only case in which a SN was clearly
detected in the field of a GRB within stringent temporal and spatial
limits \cite{gal}; the discovery and monitoring of optical polarization
in the OTs of GRB990510 \cite{wij} and GRB990712 \cite{rol}; the
determination of the redshift of GRBs 990510, 990712 \cite{pmv00} and
991216 \cite{pmv01} using VLT-Antu; the NTT detection of two very red
afterglows associated with GRB980329 \cite{pal} and GRB990705 \cite{mas},
their color being most likely due to high local absorption in the host
galaxy or to high redshift; and the discovery of the farthest GRB observed
so far, located at redshift $z$ = 4.5 \cite{and}.
Also, in several cases the detection and the spectrophotometric
observations of host galaxies associated with GRBs were accomplished with
VLT-Antu and NTT ESO telescopes.

\vspace{-.2cm}

\section{The ESO Large Programme}

\vspace{-.2cm}

In the fall of 1999, the Italian-Dutch collaboration was extended to
three more European countries, i.e. Denmark, Germany and Spain, where
groups committed to the search and followup of GRB OTs were operating and
on several occasions had already worked together with the Italian and
Dutch teams.

This led to the formation of a Consortium and to the submission of a joint
ESO proposal for a Large Programme (LP) of GRB follow-up, spanning 2
years. This programme, formerly conceived and fostered by Jan van
Paradijs, has been approved and became effective on April 1, 2000
(start of Period 65), with Ed van den Heuvel as Principal Investigator
(PI).

In order to exploit the high detection rate and the unprecedented
localization capabilities of HETE-II, which will allow increasing
the number of accurate GRB locations observable from ESO, the Large
Programme has been structured in order to achieve the following aims:  
(i) determining the redshift distributions of GRBs; this programme may
increase the redshift sample by a factor of two; (ii) studying the nature
of circumsource environment, and probing the `dark' afterglow population; 
(iii) establishing the reality and nature of GRB-supernova connection;
(iv) constraining the physics of the `fireball' model; (v) performing the
first-ever follow-up observations of the subclass of short-duration GRBs. 

In the European Collaboration each of the 5 nodes has a contact person, or
`captain', who acts as a link among his/her node, the PI and the other
captains. These persons are: Alberto Castro-Tirado (Spanish node), Jochen
Greiner (German node), Jens Hjorth (Danish node), Elena Pian (Italian
node) and Paul Vreeswijk (Dutch node).

Besides the ESO LP, an informal agreement among the nodes of the European
Consortium also exists for collaborating at other (non-ESO) telescopes,
mainly located in the Northern Hemisphere and at which each single node
has an approved GRB ToO proposal. This is to guarantee a coverage as 
complete as possible in terms of wavelength and of temporal baseline for
each single GRB. 

Concerning the guidelines for the activation of ESO telescopes, it was
agreed that the 5 nodes rotate so that the node on duty activates the ToO
request soon after the notification of a fast and precise GRB localization
and after having consulted the PI and the captains of the other nodes.
Each node on duty is supported by a `backup' node in case of need (e.g.
when extra manpower is required due to a very large and concentrated data
flow).

As a rotation criterion, it was first decided that the turn of the ESO ToO
activation was shifted from one node to the following after having
detected an OT with VLT; then the policy was changed so that, effective
November 1, 2000, each node is on duty for 15 days.
Then, the turn passes to the next node in line. Again, the node on duty
has the direct responsibility of the data reduction, analysis and
publication. In December 2000 the European Consortium was extended to a
sixth, `North-Atlantic', group (with Andy Fruchter as captain) formed by
astronomers from US and British institutes who were already included in
the LP as co-investigators.

\vspace{-.2cm}

\section{Results from ESO LP}

\vspace{-.2cm}

Up to the end of December 2000, ESO LP observations were activated 9
times. In 5 cases (GRBs 000528, 000529, 000607, 000801 and 001204) no
afterglow was detected in optical or NIR; in one case (GRB000830) a
faint and very red fading object was discovered, but its nature still
needs confirmation; in three cases (GRBs 000911, 001007 and 001011) an OT
was detected and monitored. Work on these objects is still in progress.

Besides the search and monitoring of OTs, the Euro-US GRB consortium is
also involved in the study of the host galaxies of GRBs detected so far
and observable from La Silla and Paranal. Indeed, up to one third of the
total time allocated to the LP is devoted to spectrophotometry of hosts,
in order to determine their main parameters (redshift and star-formation
rate) and to construct their optical/NIR broadband spectral energy
distributions.

%

\end{document}